\documentclass[aps,prd,showpacs,twocolumn,amsmath,10pt,superscriptaddress,floatfix,nofootinbib]{revtex4-1}

\usepackage{epsfig,amssymb,amsfonts,amsmath,bm,gensymb,color}

\newcommand{\be}{\begin{equation}}
\newcommand{\ee}{\end{equation}}
\newcommand{\bea}{\begin{eqnarray}}
\newcommand{\eea}{\end{eqnarray}}
\newcommand{\beas}{\begin{eqnarray*}}
\newcommand{\eeas}{\end{eqnarray*}}
\newcommand{\hhh}{$\vphantom{\Bigl[}$}
\newcommand{\alphaQq}{\alpha_S^{\bar{Q}q}}
\newcommand{\alphacq}{\alpha_S^{\bar{c}q}}
\newcommand{\alphabq}{\alpha_S^{\bar{b}q}}
\newcommand{\alphacc}{\alpha_S^{\bar{c}c}}
\newcommand{\alphabb}{\alpha_S^{\bar{b}b}}
\newcommand{\Ccq}{C_0^{\bar{c}q}}
\newcommand{\Cbq}{C_0^{\bar{b}q}}
\newcommand{\Ccc}{C_0^{\bar{c}c}}
\newcommand{\Cbb}{C_0^{\bar{b}b}}
\newcommand{\ds}{\displaystyle}
\graphicspath{{Figs.dir/}}

\begin{document}

\title{QCD string in excited heavy-light mesons and heavy-quark hybrids}

\author{Yu.S. Kalashnikova}
\affiliation{Institute for Theoretical and Experimental Physics, 117218, B.Cheremushkinskaya 25, Moscow, Russia}
\affiliation{National Research Nuclear University MEPhI, 115409, Kashirskoe highway 31, Moscow, Russia}

\author{A.V. Nefediev}
\affiliation{Institute for Theoretical and Experimental Physics, 117218, B.Cheremushkinskaya 25, Moscow, Russia}
\affiliation{National Research Nuclear University MEPhI, 115409, Kashirskoe highway 31, Moscow, Russia}
\affiliation{Moscow Institute of Physics and Technology, 141700, Institutsky lane 9, Dolgoprudny, Moscow Region, Russia}

\begin{abstract}
The QCD string model is employed to evaluate the masses of orbitally and radially excited heavy-light mesons and lightest hybrids in the spectrum
of charmonium and bottomonium. The number of parameters of the model is reduced to only seven which are the string tension, the two values of the
strong coupling constant (one for heavy-light and $\bar{c}c$ mesons and one for $\bar{b}b$ mesons), and the four overall spectrum shift constants
which depend on the quark contents of the particular meson or hybrid family. A few well-established states in the spectrum of heavy-light and
heavy-heavy mesons
are used to fix these parameters, and then the masses of other mesons and hybrids come out as predictions of the model which are
confronted with the existing experimental data, and a few suggestions are made concerning yet not measured quantum numbers of some states in the
spectrum of charmonium and bottomonium.
\end{abstract}

\pacs{11.30.Rd, 12.38.Aw, 14.40.-n}

\maketitle

\section{Introduction}

In the last decade, hadronic physics of heavy flavours has experienced a renaissance due to numerous discoveries made in various experiments. In
particular, $B$-factories at $e^+e^-$ colliders and the LHC play an especially important role in this process.
While $B$-factories typically operate at the energies around the $\Upsilon(4S)$ bottomonium, they have a potential to scan the
region of higher energies, too. Specifically, studies around the $\Upsilon(10860)$ resonance, which is conventionally identified as the $\Upsilon(5S)$
bottomonium, revealed many new and intriguing features---see, for example, reviews \cite{Brambilla:2010cs,Drutskoy:2012gt,Brambilla:2014jmp}. Indeed,
at the energies around 11~GeV, a
few new bottom thresholds are open. For example, studies in the vicinities of the thresholds $B^{(*)}\bar{B}^*$
allowed the Belle Collaboration to discover the charged $Z_b$ bottomoniumlike resonances \cite{Belle:2011aa}, which now attract a lot of attention due
to their exotic
nature. It still remains an open question whether or not the region near the next vector bottomonium, $\Upsilon(11020)$, can also be reached for
systematic studies by Belle-II but, in any case, additional theoretical information about this region is of paramount
importance for the field. For example, an unambiguous identification of the nature of the $\Upsilon(11020)$ resonance
and establishing the exact position of the higher-lying open-bottom thresholds are important tasks for future experiments, especially for the
$B$-factories of the new generation, like Belle-II. In particular, this amounts to making reliable predictions for the masses of excited
heavy-light $B$ mesons as well as for bottomonium hybrids. Meanwhile, the current situation with the spectroscopy of these states looks somewhat
ambiguous. From the theory
side, in the spectrum of heavy-light mesons containing a heavy quark $Q$, there should exist a positive-parity quadruplet of states $(0^+, 1^+, 1^+,
2^+)$ which in the quark-model language corresponds to
$P$-level quarkonia. The heavy-quark symmetry (exact in the limit $m_Q\to\infty$) implies a particular splitting pattern within this quadruplet and
leads to the formation of two doublets, $(0^+,1^+)$ and $(1^+,2^+)$, with a fixed value of the light-quark total momentum, $j_q=1/2$
and $j_q=3/2$, respectively. Mass degeneracy within each doublet, exact in the limit $m_Q\to\infty$, is
removed for a finite heavy-quark mass, so that the actual splitting pattern between the $P$-level heavy-light mesons may differ substantially from
that in the
strict heavy-quark limit. All members of the quadruplet in the spectrum of $D$ mesons are known experimentally (see Table~\ref{tab:cq}), while
the situation with similar states in the spectrum of $B$ mesons is more uncertain since only two states of four are unambiguously identified,
and there exists a candidate for the third state (see Table~\ref{tab:bq}). In addition, a few more candidate states in the spectrum of $D$ and $B$
mesons exist---see Refs.~\cite{Agashe:2014kda,Aaltonen:2013atp,Aaij:2015qla}---the quantum numbers of which are not yet identified. 
Identification of these
states and predictions for not yet observed ones is a challenge for phenomenologists.

Another intriguing prediction of QCD is the existence of mesons with an excited gluonic degree of freedom---the so-called hybrids.
So far, there is no clear experimental signal of the existence of hybrid mesons; however candidates do appear from time to time. For example, the 
state
$Y(4260)$ \cite{Aubert:2005rm} demonstrates some feature expected from a charmonium hybrid; namely, it has the mass close to the lattice
predictions for such a hybrid \cite{Luo:2005zg}, and, what is more important, it has a decay pattern (small electronic width and
not seen open-charm decays of a particular type) that is not typical for conventional mesons but is specific for hybrids
\cite{LeYaouanc:1984gh,Iddir:1998yc,Kalashnikova:1993xb,Kou:2005gt,Isgur:1985vy,Close:1994hc}. However, further studies of the open-charm decays
of this state \cite{Pakhlova:2009jv} do not confirm its hybrid nature. Discussion of alternative models for $Y(4260)$ can be found in
Ref.~\cite{Wang:2013kra}.

There exists a vast literature on hybrids, so let us mention some of many relevant references. For example, results of lattice simulations are
reported in
Refs.~\cite{Juge:1999ie,Michael:2003xg,Liu:2005rc,Luo:2005zg,Burch:2007fj,Dudek:2007wv,Dudek:2008sz}, and predictions of various models can be
found in Refs.~\cite{Chanowitz:1982qj,Barnes:1982tx} (bag model), Ref.~\cite{Barnes:1995hc} (flux-tube model),
Refs.~\cite{Cotanch:2001mc,General:2006ed,LlanesEstrada:2000hj} (Coulomb-gauge QCD approach), Ref.~\cite{Abreu:2005uw} (potential quark model),
Refs.~\cite{Horn:1977rq,LeYaouanc:1984gh,Iddir:1998yc} (constituent gluon model), and
Refs.~\cite{Simonov:2001rn,Kalashnikova:1995zw,Simonov:2004rh,Kalashnikova:2002tg,Kalashnikova:2008qr,Buisseret:2006wc,Buisseret:2007ed}
(QCD string approach).

In this paper, the QCD string approach is used to provide a self-consistent description of heavy-light radially and orbitally excited $D$ and $B$
mesons, low-lying heavy
$\bar{c}c$ and $\bar{b}b$ mesons, and the lowest $\bar{c}cg$ and $\bar{b}bg$ hybrids. Parameters of the corresponding Hamiltonians are totally
fixed from the masses of a few well-established heavy-light and heavy-heavy mesons. Then, the masses of other heavy-light mesons, including
radially excited as well as $P$- and $D$-wave ones, come out as predictions. Also, in the given approach, the lowest vector bottomonium hybrid is
predicted to possess the mass around 11.04 GeV that places
it just in the vicinity of the $B^{(*)}B_J$ thresholds, with $B_J$ ($J=0,1,2$) denoting the quadruplet of positive-parity $B$-mesons. Constraints
from the
heavy-quark spin symmetry which suppress decays for a genuine vector bottomonium to the corresponding open-bottom channels \cite{Li:2013yka} may give
us a clue to understanding the
nature of the $\Upsilon(11020)$ resonance.
The results obtained emphasise the importance of studies of the energy region around 11 GeV in the future high-statistics and high-precision
experiments and, in particular, are expected to be relevant for the physical programme of the next-generation $B$-factories.

\section{Hamiltonians of mesons and hybrids}

The QCD string model has a long history. It is based on the Vacuum Background Correlators Method (see review \cite{DiGiacomo:2000irz} and references
therein), and its
application to the simplest hadronic system---the quark-antiquark meson---can be found in Refs.~\cite{Dubin:1994vn,Dubin:1995vw}. A complementary
approach which radically simplifies the algebra related to the relativistic kinematics is the einbein field formalism \cite{Brink:1976uf}. It
allows one to reduce the fully relativistic kinematics to an effectively nonrelativistic one with the help of auxiliary degrees of
freedom
provided by the einbeins. The details of the formalism and relevant references can be found in Ref.~\cite{Kalashnikova:1996pu}. If
einbeins are treated as variational parameters, the suggested approach is applicable to a wide class of hadronic systems, including hybrids and
glueballs \cite{Dubin:1994vn,Simonov:1993da,Kalashnikova:1995zw,Kaidalov:1999de}. A detailed discussion of the variational procedure based on the
einebin field approach can be found in Ref.~\cite{Kalashnikova:2001ig}.

In the QCD string approach, the Hamiltonian of a hadron can be written in the form
\be
H=H_0+V_{\rm str}+V_{\rm SD},
\label{Hdec}
\ee
where $H_0$ describes the dynamics of the spinless quarks interacting through the linear-plus-Coulomb potential, $V_{\rm str}$ is the string
correction which accounts for the proper dynamics of the QCD
string \cite{Dubin:1994vn}, and $V_{\rm SD}$ describes spin-dependent interactions. In particular, for the quark-anti\-quark meson in its
centre-of-mass frame one has
\cite{Simonov:1999qj,Kalashnikova:2000tg,Kalashnikova:2001ig,Kalashnikova:2001px,Kalashnikova:2005rm}
\be
H_0=\sum_{i=1}^2\left(\frac{{\bm p}^2+m_i^2}{2\mu_i}+\frac{\mu_i}{2}\right)+\sigma
r+V_{\rm Coul}-C_0,\label{H0}
\ee
\be
V_{\rm Coul}=-\frac43\frac{\alpha_S}{r},\quad V_{\rm str}=-\frac{\sigma (\mu_1^2+\mu_2^2-\mu_1\mu_2)}{6\mu_1^2\mu_2^2}\frac{{\bm L}^2}{r},
\label{Vstr}
\ee
\be
V_{\rm SD}=V_{\rm LS}+V_{\rm SS}+V_{\rm ST},
\ee
where the subscripts LS, SS, and ST stand for the spin-orbital, hyperfine, and spin-tensor interaction, respectively (for the details see, for
example, Ref.~\cite{Kalashnikova:2001px}). The constant $C_0$ provides an overall shift of the spectrum. The quantities $\mu_{1,2}$ are
the einbein fields interpreted as dynamical masses of the quarks. For each particular eigenstate of Hamiltonian (\ref{Hdec}) their values are found
from the requirement that the corresponding eigenenergy takes an extremal value.

Similarly, for a hybrid meson containing two quarks and a gluon one has
\cite{Kalashnikova:2008qr,Kalashnikova:2009zz}
\bea
H_0&=&\frac{\mu_q+\mu_{\bar{q}}+\mu_g}{2}+\frac{m_q^2+{\bm p}_q^2}{2\mu_q}+
\frac{m_{\bar q}^2+{\bm p}_{\bar q}^2}{2\mu_{\bar q}}+\frac{{\bm p}_g^2}{2\mu_g}\nonumber\\
&+&\sigma |{\bm r}_q-{\bm r}_g|
+\sigma |{\bm r}_{\bar q}-{\bm r}_g|+V_{\rm Coul}-C_0,\label{h0}\\
V_{\rm Coul}&=&-\frac{3\alpha_s}{2|{\bm r}_q-{\bm r}_g|}-\frac{3\alpha_s}
{2|{\bm r}_{\bar q}-{\bm r}_g|}+\frac{\alpha_s}{6|{\bm r}_q-{\bm r}_{\bar q}|},\nonumber
\eea
$$
V_{\rm SD}=V_{\rm LS}^{(q\bar{q})}+V_{\rm LS}^{(g)}+V_{\rm SS}+V_{\rm ST}^{(q\bar{q})}+V_{\rm ST}^{(g)},
\label{sd}
$$
where $V_{\rm Coul}$ describes the pairwise colour Coulomb interactions \cite{Horn:1977rq}, and the string correction
(not quoted here) depends on
the angular momenta between the quarks and the gluon. For hybrids with the quark and the antiquark of the same flavour one can set $m_q=m_{\bar q}=m$, so that $\mu_q=\mu_{\bar
q}=\mu$, and the centre-of-mass motion in this three-body system can be separated with the help of the standard Jacobi coordinates,
\be
{\bm r}={\bm r}_q-{\bm r}_{\bar{q}},\quad
{\bm \rho}={\bm r}_g-\frac{\mu_q{\bm r}_q+\mu_{\bar{q}}{\bm r}_{\bar{q}}}{\mu_q+\mu_{\bar q}}=
{\bm r}_g-\frac{{\bm r}_q+{\bm r}_{\bar{q}}}{2},
\label{jacobi}
\ee
defined in terms of the effective dynamical masses of the quarks. For the explicit form of the Hamiltonian used in the calculations and for further
details see Ref.~\cite{Kalashnikova:2008qr}.

Due to the presence of extra degrees of freedom, hybrids possess properties severely different from the properties of conventional quark-antiquark
mesons. In particular, while quantum numbers of the latter follow the standard scheme, $P=(-1)^{l_{q\bar{q}}+1}$ and
$C=(-1)^{l_{q\bar{q}}+s_{q\bar{q}}}$, so that exotic quantum numbers $1^{-+}$ are not accessible, for the one-gluon hybrid one can find that
(see Ref.~\cite{Kalashnikova:2008qr} and references therein)
\be
P=(-1)^{l_{q\bar{q}}+j},\quad C=(-1)^{l_{q\bar{q}}+s_{q\bar{q}}+1},
\label{JPCmag}
\ee
for the magnetic hybrid ($l_g=j$) and
\be
P=(-1)^{l_{q\bar{q}}+j+1},\quad C=(-1)^{l_{q\bar{q}}+s_{q\bar{q}}+1},
\label{JPCel}
\ee
for the electric hybrid ($l_g=j \pm 1$), where $l_g$ is the angular momentum of the gluon relative to the quark-antiquark pair, $j$ is
the total momentum of the gluon, and $l_{q\bar{q}}$,  $s_{q\bar{q}}$ are the angular momentum and the spin of the quark-antiquark system,
respectively. So, the given quantum numbers can be achieved both for electric and magnetic hybrids.

Notice that electric hybrids possess such a large decay width into two $S$-wave heavy-light mesons that they can hardly be observed
\cite{Iddir:1998yc}. The situation for the magnetic hybrid is opposite because such a decay is forbidden for it by a well-known selection rule
\cite{LeYaouanc:1984gh,Iddir:1998yc,Kalashnikova:1993xb,Kou:2005gt,Isgur:1985vy,Close:1994hc,Isgur:1985vy}. Then, while decays into one $S$-wave and
one $P$-wave meson with open flavour are allowed, the corresponding widths are relatively small. Thus, in what follows, only lowest magnetic hybrids 
will be
considered, namely the vector $1^{--}$ one with
\be
s_{q\bar{q}}=0,\quad l_{q\bar{q}}=0,\quad l_g=1,\quad j=1
\ee
and three $C$-even $J^{-+}$ ($J=0,1,2$) siblings with
\be
s_{q\bar{q}}=1,\quad l_{q\bar{q}}=0,\quad l_g=1,\quad j=1.
\ee

\section{Parameters and procedure}

\begin{table*}[t]
\begin{center}
\begin{tabular}{|c|c|c|c|c|c|c|}
\hline
$^{2S+1}L_J$ (HQ term)          & Meson  & $J^P$  & Mass (theor.), MeV  & Width (theor.)&   Mass (exp.), MeV  &\hhh Width (exp.), MeV      \\
\hline
$^3P_0(P_{1/2})$  & $D_0(2400)$  & $0^+$     & 2343  & broad        & $2318\pm 29/2403\pm 40$    & \hhh $267\pm 40/283\pm 40$         \\
\hline
$P_1^l(P_{1/2}\cos\theta_D-P_{3/2}\sin\theta_D)$  & $D_1(2420)$  & $1^+$     & 2423   & narrow        & $2421.4\pm 0.6/2423.2\pm 2.4$ & \hhh
$27.4\pm 2.5/25\pm 6$ \\
\hline
$P_1^h(P_{1/2}\sin\theta_D+P_{3/2}\cos\theta_D)$  & $D_1(2430)$  & $1^+$     & 2441   & broad        & $2427\pm 40/$---    & \hhh
$384^{+130}_{-110}/$--- \\
\hline
$^3P_2(P_{3/2})$  & $D_2(2460)$  & $2^+$     & 2463     & narrow      & $2462.6\pm 0.6/2464.3\pm 1.6$ & \hhh $49.0\pm 1.3/37\pm 6$           \\
\hline
\end{tabular}
\end{center}
\caption{Masses of $P$-level $D$ mesons. $P_{1/2}$ and $P_{3/2}$ indicate the heavy-quark (HQ) states doubly degenerate in the strict limit
$m_c\to\infty$. The mixing angle is
$\theta_D\approx 60\degree$. Experimental data are taken from the live update of PDG \cite{Agashe:2014kda} and are quoted as
$M(D_J^0)/M(D_J^{\pm})$.}\label{tab:cq}
\end{table*}

\begin{table*}[t]
\begin{center}
\begin{tabular}{|c|c|c|c|c|c|c|}
\hline
$^{2S+1}L_J$ (HQ term)       & Meson  & $J^P$  & Mass (theor.), MeV  & Width (theor.) & Mass (exp.), MeV  &\hhh Width (exp.), MeV      \\
\hline
$^3P_0(P_{1/2})$  & $B_J^*(5732)$  & $0^+$     & 5669   & broad        & $5698\pm 8$(?)      & \hhh $128\pm 18$(?)         \\
\hline
$P_1^l(P_{1/2}\cos\theta_B-P_{3/2}\sin\theta_B)$  & $B_1$  & $1^+$     & 5713   & broad        &   ---         & \hhh ---   \\
\hline
$P_1^h(P_{1/2}\sin\theta_B+P_{3/2}\cos\theta_B)$  & $B_1(5721)$  & $1^+$     & 5724  & narrow         & $5724.9\pm 2.4/5726.8^{+3.2}_{-4.0}$  & \hhh
$23\pm 5/49^{+12}_{-16}$               \\
\hline
$^3P_2(P_{3/2})$  & $B_2^*(5747)$  & $2^+$     & 5741 & narrow          & $5739\pm 5/5736.9^{+1.3}_{-1.6}$      & \hhh $22\pm 5/11\pm 5$     \\
\hline
\end{tabular}
\end{center}
\caption{Masses of $P$-level $B$ mesons. $P_{1/2}$ and $P_{3/2}$ indicate the heavy-quark (HQ) states doubly degenerate in the strict limit
$m_b\to\infty$. The mixing angle is
$\theta_B\approx 24\degree$.
Experimental data are taken from the live update of PDG \cite{Agashe:2014kda} and are quoted as
$M(B_J^0)/M(B_J^{\pm})$. State $B_J^*(5732)$, not yet confirmed and therefore tagged with the question mark, is placed
in the most appropriate cell according to its mass and width quoted in PDG \cite{Agashe:2014kda}.}\label{tab:bq}
\end{table*}

\begin{table*}[t]
\begin{center}
\begin{tabular}{|c|c|c|c|c|c|c|}
\hline
Term                 &  $2^1S_0$    &  $2^3S_1$  &  $1^3D_3$    & $1D_2^l$ & $1D_2^h$     & $1^3D_1$    \\
\hline
$J^P$                &  $0^-$       &  $1^-$     &  $3^-$       & $2^-$    & $2^-$        & $1^-$       \\
\hline
Mass (theor.), MeV   &  2532        &  2697      &  2682        & 2693     & 2794         & 2811    \\
\hline
Mass (exp.), MeV     &  $2539\pm 8$ & $2612\pm 6$& $2637\pm 6$  & ---      & $2761\pm 5$  &  ---\\
\hline
Hypothesis ${\cal D}_1$ (51)& $D(2550)$    & $D(2600)$  & $D(2640)$    & ---      & $D(2750)$    &  --- \\
\hline
Hypothesis ${\cal D}_2$ (54)& $D(2550)$    & $D(2600)$  & $D(2640)$    & ---      &  ---         &  $D(2750)$ \\
\hline
\end{tabular}
\end{center}
\caption{Masses of radially and orbitally excited $D$ mesons predicted by the model and their possible identification with
experimentally observed states taken from Ref.~\cite{Agashe:2014kda}. The number in parentheses in the hypothesis name gives the mean quadratic
deviation (in MeV) of the theoretical predictions from the experimental masses.}\label{tab:cq2}
\end{table*}

\begin{table*}[t]
\begin{center}
\begin{tabular}{|c|c|c|c|c|c|c|}
\hline
Term               &  $2^1S_0$    &  $2^3S_1$  &  $1^3D_3$ & $1D_2^l$    &   $1D_2^h$ & $1^3D_1$  \\
\hline
$J^P$              &  $0^-$       &  $1^-$     &  $1^-$    & $2^-$       &   $2^-$    & $3^-$ \\
\hline
Spin-parity type   &  UN          &  N         &  N        & UN          &   UN       & N    \\
\hline
Mass (theor.), MeV &  5853        &  5942      &  5961     & 5962        &   6061     & 6064 \\
\hline
Hypothesis ${\cal B}_1$ (80)   & ---          & $B(5970)$  & ---         & $B_J(5840)$ &  $B_J(5960)$ & ---   \\
\hline
Hypothesis ${\cal B}_2$ (79)  & ---          & ---        & $B(5970)$   & $B_J(5840)$ &  $B_J(5960)$ & ---   \\
\hline
Hypothesis ${\cal B}_3$ (36)  & $B_J(5840)$    & $B(5970)$  & ---         & ---       &  $B_J(5960)$ & ---   \\
\hline
Hypothesis ${\cal B}_4$ (32) & $B_J(5840)$    & ---        & $B(5970)$   & ---       &  $B_J(5960)$ & ---   \\
\hline
Hypothesis ${\cal B}_5$ (63)  & ---          & $B(5970)$  & ---         & $B_J(5840)$ &  ---       & $B_J(5960)$   \\
\hline
Hypothesis ${\cal B}_6$ (61) & ---          & ---        & $B(5970)$   & $B_J(5840)$ &  ---       & $B_J(5960)$   \\
\hline
\end{tabular}
\end{center}
\caption{Masses of radially and orbitally excited $B$ mesons predicted by the model and their possible identification with
experimentally observed states taken from Refs.~\cite{Agashe:2014kda} and \cite{Aaij:2015qla}. The number in parentheses in the hypothesis name gives
the mean quadratic deviation (in MeV) of the theoretical predictions
from the experimental masses. The spin-parity scheme corresponds to the one used in Ref.~\cite{Aaij:2015qla}: natural (N) spin-parity
implies that $P=(-1)^J$ while unnatural (UN) spin-parity implies that $P=(-1)^{J+1}$.}\label{tab:bq2}
\end{table*}

\begin{table*}[t]
\begin{center}
\begin{tabular}{|c|c|c|c|c|c|c|}
\hline
Meson     \hhh      &$\eta_c(1S)$&$J/\psi(1S)$  &$h_c(1P)$&$\chi_{c_1}(1P)$&$\chi_{c_0}(1P)$&$\chi_{c_2}(1P)$\\
\hline
$J^P$  \hhh      &$0^{-+}$    &$1^{--}$      &$1^{+-}$ &$1^{++}$        &$0^{++}$        &$2^{++}$\\
\hline
$^{2S+1}L_J$  \hhh  &${}^1S_0$&${}^3S_1$&${}^1P_1$&${}^3P_1$&${}^3P_0$&${}^3P_2$\\
\hline
Exp., MeV \cite{Agashe:2014kda} \hhh&2984&3097&3525&3511&3415&3556\\
\hline
Theor., MeV     \hhh      &2981&3104&3528&3514&3449&3552\\
\hline
\end{tabular}
\end{center}
\caption{Masses of the low-lying $S$- and $P$-wave $\bar{c}c$ mesons.}\label{tab:cc}
\begin{center}
\begin{tabular}{|c|c|c|c|c|c|c|}
\hline
Meson \hhh&$\eta_b(1S)$&$\Upsilon(1S)$&$h_b(1P)$&$\chi_{b_1}(1P)$&$\chi_{b_0}(1P)$&$\chi_{b_2}(1P)$\\
\hline
$J^P$ \hhh&$0^{-+}$&$1^{--}$&$1^{+-}$&$1^{++}$&$0^{++}$&$2^{++}$\\
\hline
$^{2S+1}L_J$ \hhh &${}^1S_0$&${}^3S_1$&${}^1P_1$&${}^3P_1$&${}^3P_0$&${}^3P_2$\\
\hline
Exp., MeV \cite{Agashe:2014kda} \hhh&9398&9460&9899&9893&9859&9912\\
\hline
Theor., MeV\hhh &9394&9459&9902&9895&9871&9912\\
\hline
\end{tabular}
\end{center}
\caption{Masses of the low-lying $S$- and $P$-wave $\bar{b}b$ mesons.}\label{tab:bb}
\end{table*}

\begin{table*}[t]
\begin{center}
\begin{tabular}{|c|c|c|c|c|c|c|c|}
\hline
Parameter \hhh&$\sigma$, GeV$^2$
&$\alphacq=\alphabq=\alphacc$&$\alphabb$&$\Ccq$, MeV&$\Cbq$, MeV&$\Ccc$, MeV&$\Cbb$, MeV\\
\hline
Extracted from fit for \hhh&$\bar{c}q$&$\bar{c}q$&$\bar{b}b$&$\bar{c}q$&$\bar{b}q$&$\bar{c}c$&$\bar{b}b$\\
\hline
Listed in Table \# \hhh&\ref{tab:cq}&\ref{tab:cq}&\ref{tab:bb}&\ref{tab:cq}&\ref{tab:bq}&\ref{tab:cc}&\ref{tab:bb}\\
\hline
Value \hhh &0.16 &0.54&0.42&330&70&369&50\\
\hline
\end{tabular}
\end{center}
\caption{Parameters of the model fixed from the fits to the data.}\label{tab:param}
\end{table*}

The standard procedure to deal with Hamiltonian (\ref{Hdec}) is to solve the corresponding Sch{\"o}dinger equation for the leading-order term $H_0$
and then to include other terms as perturbations. Further details can be found in
Refs.~\cite{Kalashnikova:2000tg,Kalashnikova:2001ig,Kalashnikova:2001px,Kalashnikova:2005rm,Kalashnikova:2008qr}.
It should be noticed that, unlike previous works, in this paper the number of parameters of the model is reduced to a minimum; in particular, the
quark masses are not treated as free parameters, and the remaining seven parameters are fixed in a self-consistent way for all
hadronic systems discussed. Also, updated experimental data are used. Thus, in what follows, the masses of the quarks take their standard pole values
evaluated in two loops \cite{Agashe:2014kda}. Since the isospin effects lie beyond the scope of this research, then for the light quark $q$ the
averaged value between the $u$ quark mass and the $d$ quark mass is used. Therefore,
\be
m_q=3.6~\mbox{MeV},~~ m_c=1.67~\mbox{GeV},~~ m_b=4.78~\mbox{GeV}.
\label{masses}
\ee

Then the set of parameters of the model is given by the string tension $\sigma$, the strong coupling constant $\alpha_s$, and the overall constant
shift of the spectrum $C_0$. The parameters $\sigma$ and
$\alpha_S$ can be somewhat adapted to a particular system under study; however, they are strongly constrained by phenomenology. In particular, the
string tension takes its standard value consistent with phenomenology---see Table~\ref{tab:param}. The situation with the strong coupling
constant is somewhat more subtle. It demonstrates a dependence on the scale which can be presented as \cite{Simonov:2010gb}
\be
\alpha_S(Q^2)=\left(b_0\ln\frac{Q^2+{\cal M}^2}{\Lambda_{\rm QCD}^2}\right)^{-1},
\label{af}
\ee
where $b_0$ is the one-loop coefficient of the $\beta$-function, $\Lambda_{\rm QCD}$ is the standard parameter of QCD, and ${\cal M}$ takes values of
the order
of 1--2 GeV---see the discussion and relevant references in Ref.~\cite{Simonov:2010gb}. It is easy to see that ${\cal M}\simeq m_c$, so that, for the
scales below $m_c$, $\alpha_S$ remains nearly constant and takes values close to the ``frozen'' limit $\alpha_S^{\rm fr}\approx 0.6$. Meanwhile,
since $m_b\simeq (3\mbox{-}4) {\cal M}$,
then it is natural to expect a smaller $\alpha_s$ in the $\bar{b}b$ bottomonia. In other words, the following hierarchy of the values of $\alpha_S$ is
expected,
\be
\alpha_S(m_q)\approx \alpha_S(m_c)>\alpha_S(m_b),
\label{alphasineq}
\ee
that implies that ($Q=c,b$)
\be
\alphaQq\approx\alphacc\simeq 0.5\mbox{-}0.6,\quad \alphabb\simeq 0.4\mbox{-}0.5.
\label{alphasest}
\ee

The constant $C_0$ is
treated as a free parameter of the model, and we take it to be the same in both quarkonium and hybrid sectors. 
As shown in Ref.~\cite{Simonov:2001iv}, this constant can 
be viewed as the quark self-energy which takes into account the bare quark mass renormalisation due to the confining background. Obviously, such 
a renormalisation is absent 
for gluons because of gauge invariance. Following this reasoning we also assume that the constant $C_0$ appears as the quark self-energy
while the gluon self-energy vanishes. 
This assumption finds a further phenomenological justification
in the calculations of the glueball spectrum in the QCD string approach~\cite{Simonov&Kaidalov}: the calculated glueball masses, with the gluon 
self-energy 
set equal to zero, agree well with the masses found on the lattice.

Therefore, the following procedure is adopted. First, the model is fully fixed and verified as follows:
\begin{itemize}
\item The spectrum of the $P$-level $D$ mesons (four states) is calculated---see Table~\ref{tab:cq}---and parameters $\sigma$, $\alphacq$, and $\Ccq$ 
are
adjusted to provide the best overall description of the experimentally observed masses. If both neutral ($m_0$) and charged ($m_\pm$) mesons are
measured, the isospin averaged value $(2m_\pm+m_0)/3$ is used in the fit.
\item The masses of the well-established $P$-level $B$ mesons (two states) are calculated with the string tension and the coupling $\alpha_S$ taking
the values found above, from the fit for the $D$-meson masses---see Table~\ref{tab:bq}---and the only free parameter, the constant $\Cbq$, is
adjusted this way. Notice that, since $C_0$ only provides the overall shift of the spectrum, then the splittings between the $B$ mesons are predictions.
\item The spectrum of the low-lying $\bar{c}c$ mesons (six states) is calculated---see Table~\ref{tab:cc}---and the constant $\Ccc$ is fixed this way.
As before, the splittings between the levels are not adjusted and come as predictions.
\item The spectrum of the low-lying $\bar{b}b$ mesons (six states) is calculated---see Table~\ref{tab:bb}---and the only remaining free parameters
of the model, $\alphabb$ and $\Cbb$, are determined.
\end{itemize}

For convenience, the values of the parameters extracted as explained above are collected in Table~\ref{tab:param}. It is worthwhile noticing that the
values of $\alpha_S$ comply quite well with relation (\ref{alphasineq}) and, in particular, fall into the ranges quoted in Eq.~(\ref{alphasest}). This provides an
additional self-consistence test for the approach.

For completeness, we quote here the values of the auxiliary parameters $\mu_1$ and $\mu_2$ as they come out from the calculations,
\bea
&&\mu_1(\bar{c}q;1P)=1781~\mbox{MeV},~ \mu_2(\bar{c}q;1P)=618~\mbox{MeV},\nonumber\\
&&\mu_1(\bar{b}q;1P)=4830~\mbox{MeV},~ \mu_2(\bar{b}q;1P)=694~\mbox{MeV},\nonumber\\[-3mm]
\label{mus}\\[-2mm]
&&\mu_1(\bar{b}q;1D)=4840~\mbox{MeV},~ \mu_2(\bar{b}q;1D)=765~\mbox{MeV},\nonumber\\
&&\mu_1(\bar{b}q;2S)=4847~\mbox{MeV},~ \mu_2(\bar{b}q;2S)=801~\mbox{MeV},\nonumber
\eea
where in parentheses we give the quark contents of the heavy-light system and its quantum numbers.

\section{Results for heavy-light mesons}

\begin{figure}[t]
\centerline{\epsfig{file=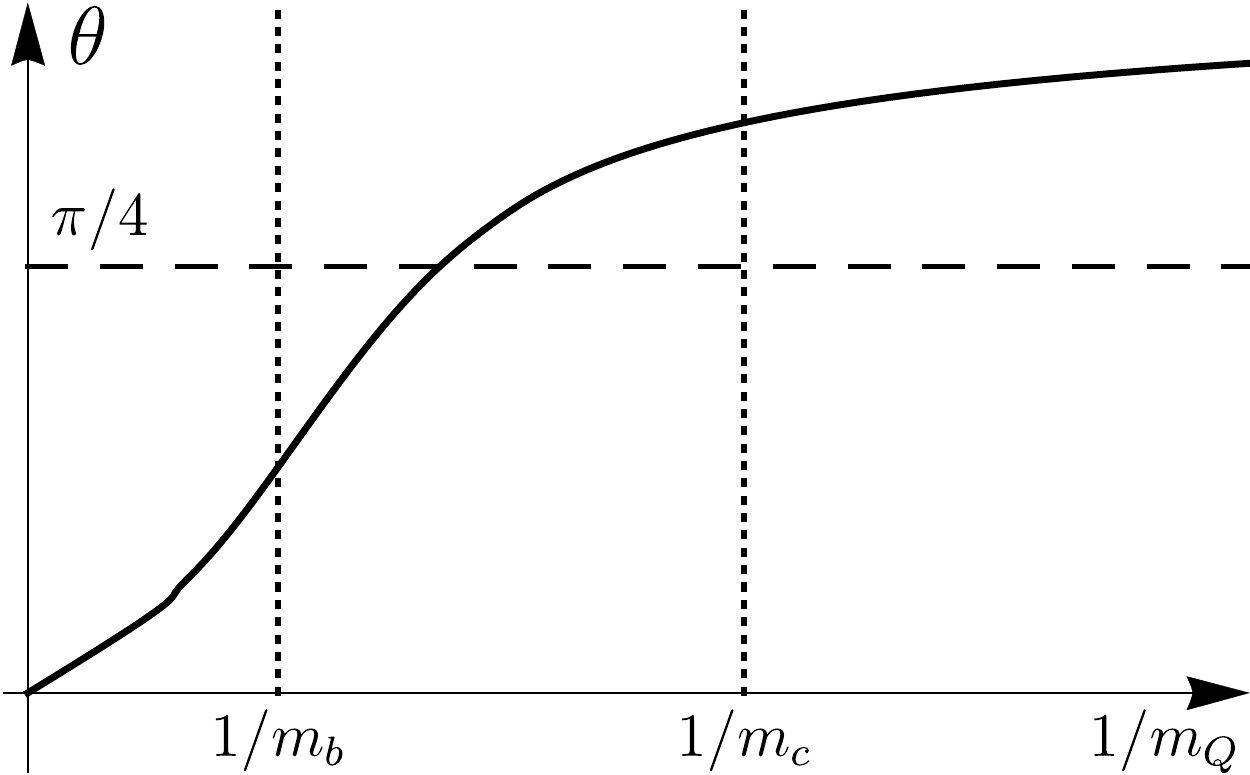, width=0.35\textwidth}}
\caption{The mixing angle $\theta$ for the $P$ levels as a function of the inversed heavy-quark
mass. The physical points for the $b$ and $c$ quarks are shown with vertical dotted lines.}\label{fig:mixing}
\end{figure}

Now, with the details of the approach described in the previous section and with the complete set of parameters of the model fixed as quoted in
Eq.~(\ref{masses}) and in Table~\ref{tab:param}, we are in a position to turn to various predictions of the model.
We start from the heavy-light $D$ and $B$ mesons---see Tables~\ref{tab:cq} and \ref{tab:bq}. As was explained above, the masses of the six
well-established experimentally states were used as input to fix the parameters of the model. From Table~\ref{tab:cq}, one can see that the
model is able to describe the spectrum of the $P$-wave $D$ mesons with a sufficiently high accuracy. The same conclusion holds for the two known
positive-parity $B$ mesons.

It has to be noticed that, in order to proceed with the identification of the heavy-light mesons, it is important to
understand the splitting
pattern in the $P$-level quadruplet. As was mentioned in the Introduction, the heavy-quark symmetry implies the formation of two degenerate
doublets, $(0^+,1^+)$ and $(1^+,2^+)$, with a fixed value of the light-quark total momentum, $j_q=1/2$ and $j_q=3/2$, respectively. Notice also that
the total quark spin is not a good quantum number in the system which does not possess $C$-parity, so that the $P$-level states with the same total
momentum but with different total spins are mixed with the spin-orbit interaction and the observed mesons appear as particular combinations of
the latter. The mixing can be parametrised through the mixing angle $\theta$ as (see Appendix~\ref{appA} for the details)
\be
\left(P_1^l\atop P_1^h\right)=
\left(
\begin{array}{cc}
\cos\theta&-\sin\theta\\
\sin\theta&\cos\theta
\end{array}
\right)
\left(P_{1/2}\atop P_{3/2}\right),
\label{thetadef}
\ee
where the superscript $l(h)$ denotes the light(heavy) member of the doublet.

The dependence of the mixing angle on the heavy-quark mass, as predicted by our model, is shown in Fig.~\ref{fig:mixing}. It is seen from the figure
that the mixing angles for the $D$ and $B$ mesons lie on different sides from the line $\theta=\pi/4$ that implies that the $P$ levels in the two
systems follow each other in a different order. Indeed, in the $D$ mesons, $\theta_D\approx 60\degree>45\degree$ and therefore the states which
are completely or predominantly given by the $P_{1/2}$ and $P_{3/2}$ levels follow one by one.
Heavy-quark symmetry constraints imply that the $P_{1/2}$ states couple to a heavy-light ground-state meson and pion in the $S$ wave while the
$P_{3/2}$ ones couple to a heavy-light ground-state
meson and pion in the $D$ wave. Thus, one expects the $P_{1/2}$ states to be broad and the
$P_{3/2}$ states to be narrow, so that the width pattern for the $D$ mesons is predicted by our model to be (broad,narrow,broad,narrow), starting from the lightest state---see
Table~\ref{tab:cq}. Conversely, for the $B$ mesons, $\theta_B\approx 24\degree<45\degree$, so that the width pattern is different, namely
(broad,broad,narrow,narrow)---see Table~\ref{tab:bq}. This makes a crucial difference between the splitting patterns of the $P$-level
$D$ and $B$ mesons.

According to this scheme, the two not yet identified members of the positive-parity quadruplet of the $B$ mesons with the quantum numbers $0^+$ and
$1^+$ are expected to be broad, with the width of the order of a few hundred MeV. Their masses are predicted to take the values around 5700 and 
5730
MeV, respectively. Then, the observed state $B_J^*(5732)$ \cite{Agashe:2014kda}, if confirmed, can be identified as the scalar meson $B_0$ which, in 
agreement with
the qualitative prediction of the model, is broad---see Table~\ref{tab:bq}.

As the next step, the masses of several radially ($n=2$) and orbitally ($l=2$) excited $D$ and $B$ mesons are calculated in the same framework and
are confronted
with the existing experimental data. The results of calculations and the hypotheses concerning a possible identification of the experimentally
observed mesons are contained in Tables~\ref{tab:cq2} and \ref{tab:bq2}. If these hypotheses are ranked according to the mean
quadratic deviation of the theoretical predictions from the experimental results (evaluated as $\Delta m=\sqrt{\sum_{n=1}^N(m_n^{\rm th}-m_n^{\rm
exp})^2/N}$, with $N$ denoting the number of states analysed, and quoted in parentheses for each hypothesis) then hypotheses ${\cal D}_1$ and
${\cal B}_4$ should be accepted as the most reliable. The details of the experimental situation with the
$B_J(5840)$ and $B_J(5960)$ candidates can be found in Ref.~\cite{Aaij:2015qla}. Our results are qualitatively compatible with similar
predictions previously made for the excited $D$ mesons in Ref.~\cite{Badalian:2011tb} (in the QCD string approach) and with those obtained recently
for the excited
$D$ and $B$ mesons in Refs.~\cite{Li:2010vx,Lu:2016bbk} (in the framework of the constituent quark model of Ref.~\cite{Lakhina:2006fy}). It
should be noticed that the masses of the excited heavy-light mesons predicted in the present work in the framework of the QCD string model typically
lie somewhat lower than those obtained in Ref.~\cite{Lu:2016bbk} that results in a slightly different suggestion for the identification of the
experimentally observed mesons with the theoretically predicted states. The origin of the discrepancy should come from the fact (i) that relativistic
dynamics is taken into account in this work while the model used in Ref.~\cite{Lu:2016bbk} is essentially nonrelativistic; (ii) that, contrary to
the purely potential approach used in Ref.~\cite{Lu:2016bbk}, the proper dynamics of the QCD string is taken into account in our model, which provides
an additional negative contribution to the energy---see Eq.~(\ref{Vstr}); and (iii) that the variational einbein field method used in this work may
somewhat overestimate the value of the wave function at the origin which governs the mass splitting between the $n^1S_0$ and $n^3S_1$ states. A detailed
comparison with other approaches and models as well as
the relevant references can be found, for example, in Refs.~\cite{Badalian:2011tb,Lu:2016bbk}.

\section{Results for heavy-quark hybrids}

We now proceed to hybrids. For the states containing the $c$ quark, the string tension and the strong coupling constant are fixed from
the spectrum of the heavy-light mesons and the constant $\Ccc$ is fixed from the spectrum of low-lying $\bar{c}c$
mesons---see Table~\ref{tab:param}.

From Table~\ref{tab:cc}, one can see
that the model describes the experimental spectrum of the $\bar{c}c$ states with high accuracy, which is especially remarkable given that only the overall shift of the spectrum
$\Ccc$ was adjusted, and all other parameters were fixed earlier. Similarly, the spectrum of the low-lying $\bar{b}b$ mesons is also
described with the same string tension while, in agreement with the discussion above, the strong coupling constant is somewhat decreased in
this case; notice
that its fitted value complies well with the estimate from Eq.~(\ref{alphasest}). The obtained values of the $\alphabb$
and $\Cbb$
are quoted in Table~\ref{tab:param}. Similarly to the $\bar{c}c$ states, the spectrum of the low-lying $\bar{b}b$ mesons is remarkably well described
by the model---see Table~\ref{tab:bb}.

To search for the eigenenergies of the Hamiltonian (\ref{H0})
we employ the variational technique described in detail in Ref.~\cite{Kalashnikova:2008qr}. In particular,
we use the Harmonic Oscillator trial wave function [$\exp(-\beta^2\mu r^2/2)$ multiplied by the appropriate spherical harmonic and 
Laguerre polynomial], that gives
\bea
&&\mu(\bar{c}c;1S)=1893~\mbox{MeV},~~\beta^2(\bar{c}c;1S)=280~\mbox{MeV},\nonumber\\
&&\mu(\bar{c}c;1P)=1866~\mbox{MeV},~~\beta^2(\bar{c}c;1P)=148~\mbox{MeV},\nonumber\\[-3mm]
\label{musbetas}\\[-2mm]
&&\mu(\bar{b}b;1S)=5019~\mbox{MeV},~~\beta^2(\bar{b}b;1S)=312~\mbox{MeV},\nonumber\\
&&\mu(\bar{b}b;1P)=4942~\mbox{MeV},~~\beta^2(\bar{b}b;1P)=128~\mbox{MeV},\nonumber
\eea
where, as before, the quark contents of the quark-antiquark system and its quantum numbers are quoted in parentheses.

With the set of the parameters from Table~\ref{tab:param} we are now in a position to predict the masses of the lowest
magnetic $\bar{c}cg$ hybrids. We use the trial wave function $\rho Y_{1m}(\hat{\rho})\exp(-\beta^2 M R^2/2)$, where $\rho$ is the 
Jakobi coordinate of the gluon relative to the centre of mass of the quark-antiquark subsystem, $R$ is the standard hyperspherical
radius defined for the three-body system $\bar{Q}Qg$, and $M=2\mu+\mu_g$---see Ref.~\cite{Kalashnikova:2008qr} for further details.
Then, the parameters $\mu$, $\mu_g$, and $\beta^2$ take the following values (in MeV):
\be
\mu(\bar{c}cg)=1778,~~\mu_g(\bar{c}cg)=1104,~~\beta^2(\bar{c}cg)=380.
\label{mbc}
\ee

The results given in Table~\ref{tab:ccg} can
be regarded as an update of the predictions contained in Ref.~\cite{Kalashnikova:2008qr}. They comply
well with the predictions found in the literature and obtained in the framework of different approaches. In particular, the bag model predicts the
mass of the lowest charm hybrid around 4 GeV \cite{Chanowitz:1982qj,Barnes:1982tx}. In the flux tube model the low-lying hybrids reside in the region
around 4.1--4.2 GeV \cite{Isgur:1984bm}. Adiabatic approximation for heavy quarks in the QCD string model in the formalism of auxiliary fields also
gives a similar result, namely $4.2 \pm 0.2$ GeV for the hybrid with the exotic quantum numbers $1^{-+}$ \cite{Buisseret:2006sz}. The mass of the tensor hybrid
is predicted to be 4.12 GeV in the potential quark model \cite{Abreu:2005uw}. Various lattice calculations also place charmonium hybrids at around
4.4 GeV \cite{Michael:2003xg,Liu:2005rc,Luo:2005zg}.

\begin{table*}[t]
\begin{center}
\begin{tabular}{|c|c|c|c|c|}
\hline
$J^P$ \hhh&$0^{-+}$&$1^{-+}$&$1^{--}$&$2^{-+}$\\
\hline
Theor., MeV \hhh&4296&4358&4430&4484\\
\hline
\end{tabular}
\end{center}
\caption{Predictions for the masses of the lowest hybrids $c\bar{c}g$.}\label{tab:ccg}
\begin{center}
\begin{tabular}{|c|c|c|c|c|}
\hline
$J^P$ \hhh&$0^{-+}$&$1^{-+}$&$1^{--}$&$2^{-+}$\\
\hline
Theor., MeV \hhh&10990&11013&11038&11057\\
\hline
\end{tabular}
\end{center}
\caption{Predictions for the masses of the lowest hybrids $b\bar{b}g$.}\label{tab:bbg}
\end{table*}

Analogously, the parameters from Table~\ref{tab:param} allow one to predict the masses of the bottomonium hybrids which are collected in
Table~\ref{tab:bbg}. The corresponding values of the parameters $\mu$, $\mu_g$, and $\beta^2$ are (in MeV)
\be
\mu(\bar{b}bg)=4813,~~\mu_g(\bar{b}bg)=1194,~~\beta^2(\bar{b}bg)=330.
\label{mbb}
\ee

It is important to notice that the vector hybrid is expected to have the mass around 11 GeV, that is it resides in the vicinity
of the $\Upsilon(11020)$ resonance. This result complies well with the predictions from the lattice which place the bottomonium hybrid at
10900(100) MeV \cite{Juge:1999ie}.

\section{Discussion and conclusions}

In this paper, we revised the QCD string approach in application to heavy-light mesons and hybrids containing heavy quarks. In contrast to earlier
works, the number of parameters is minimised and the same set of parameters, consistent with phenomenology, is used to describe simultaneously masses
of the radially and orbitally excited $D$ and $B$ mesons, low-lying $\bar{c}c$ and $\bar{b}b$ $S$-wave and $P$-wave mesons, and the lowest magnetic
$\bar{c}cg$ and $\bar{b}bg$ hybrids. The approach used in this work, on one hand being rather simple and physically transparent, on the other hand demonstrates a high
accuracy, and thus its predictions for yet not observed or not confirmed
states can be regarded as rather reliable. In particular, the $B_1(5721)$, $B_J^*(5732)$, and $B_2^*(5747)$ mesons are identified as the axial vector
($1^+$), the scalar ($0^+$), and the tensor ($2^+$) members of the $P$-level $J^+$ ($J=0,1,2$) quadruplet, respectively. The last remaining member of
the same
quadruplet is predicted to be broad and to possess the mass around 5713 MeV. Also, the states $B_J(5840)$ and $B_J(5960)$ reported recently by the
LHCb Collaboration are most probably the $2^1S_0$ and the $1D_2^h$ (here $h$ stands for the heavy member of the $1D_2$ doublet), respectively.
Finally, in the same scheme, the CDF meson $B(5970)$ can be identified with the $1^3D_3$ state (such an identification was 
also suggested in Ref.~\cite{Xiao:2014ura}). Meanwhile, we agree with the conclusion of
Ref.~\cite{Lu:2016bbk} that other hypotheses for these states should be considered seriously too, and that additional important, probably
decisive, information should be provided by the data on the decay modes of the states under study.

Finally, the masses of the lowest magnetic charmonium and bottomonium hybrids are calculated in the same model and with the parameters previously
fixed from the spectrum of ordinary mesons. Interestingly, the vector bottomonium
hybrid is predicted to have the mass of approximately 11.04 GeV that is very close to the mass of the $\Upsilon(11020)$ resonance. This may
imply a considerable admixture of the hybrid component in its wave function, in addition to the $\bar{b}b$ component
which can be identified with the radially excited $\Upsilon(6S)$ genuine $\bar{b}b$ quarkonium.

Identification of the positive-parity $B_J$ mesons given in Table~\ref{tab:bq}, together with the well-established masses of the
pseudoscalar $B$ meson and the vector $B^*$ meson \cite{Agashe:2014kda},
\be
m_B=5279~\mbox{MeV},\quad m_{B^*}=5325~\mbox{MeV},
\ee
allows one to estimate the positions of the lowest open-bottom thresholds with the $B_J$ family mesons involved,
\bea
&M(B\bar{B}_1(5721))=11005~\mbox{MeV},&\nonumber\\
&M(B^*\bar{B}_1(5721))=11050~\mbox{MeV},&\label{BB}\\
&M(B^*\bar{B}_2^*(5747))=11064~\mbox{MeV},&\nonumber
\eea
where only the narrow $B_J$ mesons are taken into account since the experimental observation of their broad partners in the open-bottom final states of
the form
(\ref{BB}) does not look feasible. The thresholds which involve two $B_J$ mesons lie considerably higher, at around 11.5~GeV.
Therefore, while the production channels for the $B_J$ family are kinematically closed for the $B$-factories working at the energies of the
$\Upsilon(4S)$
and $\Upsilon(10860)$ vector resonances, they could be observed at Belle-II in the decays of $\Upsilon(11020)$. This possibility requires an additional study
though. Since the broad members of the positive-parity quadruplet not considered here originate from the $P_{1/2}$ heavy-quark state, we concentrate
on the $P_{3/2}$ term. It has to be noticed then that production of a heavy-light meson from the $P_{3/2}$ state accompanied by a $S$-wave
$B^{(*)}$ meson---see Eq.~(\ref{BB})---is only possible if the produced light-quark pair has the total momentum equal to 1. This condition is not 
fulfilled
for the vector bottomonium where $j_{q\bar{q}}=0$, and therefore the amplitude for its decay into the $B^{(*)}B(P_{3/2})$ pair
is suppressed in the heavy-quark limit \cite{Li:2013yka}, which is certainly a good approximation for the $b$ quark.
Meanwhile, open-flavour decays of a $\bar{b}bg$ hybrid proceed through the gluon conversion into a light quark-antiquark pair which therefore carries
the quantum numbers of the vector, in particular, $j_{q\bar{q}}=1$. This implies that there is no suppression for the amplitude of the vector hybrid
decay into a pair of one $S$-wave and one $P$-wave open-bottom meson---see Ref.~\cite{Kalashnikova:2008qr,Kalashnikova:2009zz} for
the corresponding recoupling coefficients. Therefore, the decays to the final states from Eq.~(\ref{BB}) [especially to the first one, with the
threshold located below the
nominal $\Upsilon(11020)$ mass] could be used as test modes for the bottomonium hybrid in the vicinity of 11~GeV. It should be noticed, however, that
this conclusion is valid only in the strict heavy-quark limit $m_b\to\infty$. For a finite $m_b$, corrections of two
types have to be taken into account. On one hand, there exist corrections to the heavy-quark spin symmetry limit which are controlled by the small
parameter $\Lambda_{\rm QCD}/m_b$ and which are expected to be quite small, too---indeed, constraints from the heavy-quark spin symmetry are
typically very well met in bottomonium systems. On the other hand, as was mentioned above, the physical meson $B_1(5721)$ is a mixture of both
$B_{1/2}$ and $B_{3/2}$ states governed by the mixing angle $\theta_B$---see Fig.~\ref{fig:mixing} and Table~\ref{tab:bb}. Therefore, the probability
of the decay $\Upsilon(11020)\to B\bar{B}_1(5721)$ is proportional to $\sin^2\theta_B$ for the $\Upsilon(11020)$ as a genuine $\bar{b}b$ quarkonium,
and it is proportional to $\cos^2\theta_B$ for the hybrid. For $\theta_B\ll 1$, this mode could have been regarded as a smoking gun for the hybrid
nature of the $\Upsilon(11020)$ resonance. Meanwhile, the actual mixing angle is $\theta_B\approx 24\degree$ that gives $\sin^2\theta_B\approx 0.17$
and
$\cos^2\theta_B\approx 0.83$. Thus, although $\sin^2\theta_B\ll \cos^2\theta_B$, it remains to be seen whether or not such a suppression factor is
sufficient to allow one to distinguish between the genuine quarkonium and the hybrid lying at around 11~GeV. However, in any case, studies of the
decays to the final states from Eq.~(\ref{BB})\footnote{Although the last two thresholds in Eq.~(\ref{BB}) formally lie above the nominal mass of the
$\Upsilon(11020)$ resonance, due to the finite width of the latter as well as the finite widths of the $B$ mesons, they might be possible to observe.}
appear to be a very interesting and promising source of information for the phenomenology of bottomonium, and therefore data taking at $B$-factories 
of
the next generation at the energies around 11 GeV and above look quite promising (see also the discussions in Ref.~\cite{Drutskoy:2012gt}). Given that
the above decays are expected to occur near their respective thresholds, the corresponding line shapes should demonstrate a typical threshold
behaviour that makes them appealing also for various studies of the threshold phenomena.

\begin{acknowledgments}
The authors would like to thank F.-K. Guo and M. Voloshin for discussions. This work is supported by the Russian Science
Foundation (Grant No. 15-12-30014).
\end{acknowledgments}

\appendix

\section{Splitting scheme for $P$-level heavy-light mesons}\label{appA}

For an arbitrary heavy-quark mass $m_Q$ the physical observed states with the quantum numbers $J^P=1^+$, conveniently denoted as $P_1^l$ and $P_1^h$
for the light and the heavy member of the doublet, respectively, are presented as particular combinations of the $\{^{2S+1}L_J\}$ basis vectors $^1P_1$
and $^3P_1$,
\be
\left(P_1^l\atop P_1^h\right)=
\left(
\begin{array}{cc}
\cos\theta(m_Q)&-\sin\theta(m_Q)\\
\sin\theta(m_Q)&\cos\theta(m_Q)
\end{array}
\right)
\left({}^1P_1\atop {}^3P_1\right).
\label{thetaP}
\ee

The mixing matrix in Eq.~(\ref{thetaP}) can be found as
\be
\left(
\begin{array}{cc}
\ds\frac{E_2^{(0)}-E_1}{\sqrt{(E_2^{(0)}-E_1)^2+V^2}} & \ds-\frac{V}{\sqrt{(E_2^{(0)}-E_1)^2+V^2}}\\[4mm]
\ds\frac{E_2^{(0)}-E_2}{\sqrt{(E_2^{(0)}-E_2)^2+V^2}} & \ds-\frac{V}{\sqrt{(E_2^{(0)}-E_2)^2+V^2}}
\end{array}
\right),
\label{sol}
\ee
where
\be
E_1^{(0)}\equiv \langle {}^1P_1|H|{}^1P_1\rangle,\quad E_2^{(0)}\equiv\langle {}^3P_1|H|{}^3P_1\rangle,
\ee
\be
V\equiv \langle {}^1P_1|H|{}^3P_1\rangle=\langle {}^3P_1|H|{}^1P_1\rangle,
\ee
and $E_1$ and $E_2$ are the solutions of the secular equation,
\be
\mbox{det}\left(
\begin{array}{ccc}
E_1^{(0)}-E& V\\
V& E_2^{(0)}-E
\end{array}
\right)=0,
\label{eq}
\ee
that is
\be
E_{1,2}=\frac12(E_2^{(0)}-E_1^{(0)})\pm\sqrt{\frac14(E_2^{(0)}-E_1^{(0)})^2+V^2}.
\ee

In the strict heavy-quark limit, $m_Q\to\infty$, the mixing matrix from Eq.~(\ref{thetaP}) takes a universal form which corresponds to the ``ideal''
mixing,
\be
\left(P_1^{l(0)}\atop P_1^{h(0)}\right)=
\left(
\begin{array}{cc}
\cos\theta(\infty)&-\sin\theta(\infty)\\
\sin\theta(\infty)&\cos\theta(\infty)
\end{array}
\right)
\left({}^1P_1\atop {}^3P_1\right),
\label{M0}
\ee
where $\cos\theta(\infty)=1/\sqrt{3}$ and $\sin\theta(\infty)=\sqrt{2/3}$.

Obviously, in the same heavy-quark limit, one can identify the physical states $P_1^{l(0)}$ and $P_1^{h(0)}$ with the states
$P_{1/2}$ and $P_{3/2}$ in heavy-quark limit, respectively.\footnote{The inversed level ordering, that is $M(P_{1/2})>M(P_{3/2})$, is also possible in the given
model---see Ref.~\cite{Kalashnikova:2001px}---however, the updated experimental data used in this work favour the direct ordering.} Then
Eqs.~(\ref{thetaP}) and (\ref{M0}) together give relation (\ref{thetadef}) between the wave functions of the physical states and the heavy-quark basis
states, with $\theta=\theta(m_Q)-\theta(\infty)$. Dependence of the angle $\theta$ from the heavy-quark mass is depicted in Fig.~\ref{fig:mixing}.

\end{document}